\documentclass[twocolumn,showpacs,preprintnumbers,superscriptaddress,amsmath,amssymb]{revtex4-1}
\usepackage{wallpaper}
\usepackage{graphicx}
\usepackage{dcolumn}
\usepackage{bm}
\usepackage{color}
\usepackage{multirow}
\usepackage{CJK}
\usepackage{mathptmx}
\usepackage[colorlinks,linkcolor=blue,anchorcolor=blue,citecolor=blue]{hyperref}

\begin{document}
\begin{CJK*}{UTF8}{gbsn}
\renewcommand\arraystretch{1.2}

\title{Single-state or low-lying-states dominance mechanism of $2\nu\beta\beta$-decay nuclear matrix elements}

\author{W. L. Lv~(吕万里)}
\address{School of Nuclear Science and Technology, Lanzhou University, Lanzhou 730000, China}
\author{Y. F. Niu~(牛一斐)}\email{niuyf@lzu.edu.cn}
\address{School of Nuclear Science and Technology, Lanzhou University, Lanzhou 730000, China}
\author{D. L. Fang~(房栋梁)}
\address{Institute of Modern Physics, Chinese Academy of Sciences, Lanzhou 730000, China}
\address{School of nuclear science and technology, University of Chinese Academy of Sciences, Beijing 100049, China}
\author{C. L. Bai~(白春林)}
\address{Department of Physics, Science, and Technology, Sichuan University, Chengdu 610065, China}

\date{\today}

\begin{abstract}
The $2\nu\beta\beta$-decay nuclear matrix elements (NMEs)
for 11 nuclei are studied with the self-consistent quasiparticle random phase approximation (QRPA) based on Skyrme Hartree-Fock-Bogoliubov (Skyrme HFB) model. As a common feature pointed out in
[\href{https://journals.aps.org/prc/abstract/10.1103/PhysRevC.98.064325}{Phys. Rev. C \textbf{98}, 064325 (2018)}],
negative contributions in the running sums of NMEs  are found,
and play important roles in the fulfillment of the single-state dominance or
low-lying-states dominance hypothesis.
By comparing the results of QRPA model and quasiparticle Tamm-Dancoff approximation (QTDA) model,
we find that the negative contributions are due to the enhanced ground-state correlations,
which are brought by the backward amplitude in QRPA model and
tuned by strong isoscalar pairing interaction.
The enhancement of ground-state correlations will change the signs of GT$^{+}$ transition amplitudes of higher-lying states and leads to the negative contributions in the running sum.
\end{abstract}

\pacs{21.60.Jz, 23.40.−s, 23.40.Hc}
\maketitle
\end{CJK*}

\section{Introduction}
\label{secIntro}

Double-$\beta$ ($\beta\beta$) decay is a rare transition process between nuclei $(A,Z)$ and $(A,Z+2)$.
There may exist two modes of this decay. One is the two-neutrino $\beta\beta$ ($2\nu\beta\beta$) decay
with the emission of two antineutrinos.
 As a second-order weak process, it was first suggested by Mayer \cite{Mayer1935}
and has been first observed for $^{82}$Se \cite{Elliott1987} in 1980's and for other 10 different nuclei later \cite{Barabash2015}.
The other mode is the so-called neutrinoless $\beta\beta$ ($0\nu\beta\beta$) decay,
which is a lepton-number-violating nuclear process
and occurs only if neutrinos are their own antiparticles, i.e., the Majorana particles \cite{Haxton1984,Avignone2008}.
$0\nu\beta\beta$ decay is the only practical way to determine whether neutrinos are Majorana particles
and is helpful to address the questions of the absolute neutrino mass scale and the neutrino mass hierarchy,
provided that the corresponding nuclear matrix elements (NME) $M^{0\nu}$ are calculated exactly
\cite{Faessler1998,Suhonen1998,Engel2017,Ejiri2019,Yao2021}.
However, the deviations of $M^{0\nu}$ from different nuclear models
vary up to a factor around three times (c.f. Figure 26 in Ref. \cite{Yao2021}).
The main reason of such a divergence
stems from the difference of nuclear many-body wave functions obtained from different nuclear models.
While $0\nu\beta\beta$ decay is related to the underlying new physics, $2\nu\beta\beta$ decay is free from the unknown neutrino properties.
Also the corresponding lepton phase-space factor has been calculated with high precision \cite{Kotila2012,Stoica2013}.
Therefore, the NME of $2\nu\beta\beta$ decay, $M^{2\nu}$, provides a test ground for nuclear structure calculations.

As a second-order weak process, the $M^{2\nu}$ of $2\nu\beta\beta$ decay
involves a summation over all the virtual intermediate states of the nucleus $(A,Z+1)$,
which connects the initial nucleus $(A,Z)$ and final nucleus $(A,Z+2)$ by the $\beta$-decay like transitions.
Because of isospin symmetry, the Fermi transition is highly suppressed so that
only the NME of Gamow-Teller (GT) transition
$M^{2\nu}_{\rm GT}$ is considered in the calculations.
In 1984, Abad proposed the single-state dominance (SSD) hypothesis, which states that the $M^{2\nu}_{\rm GT}$ is dominated
by the virtual transitions through the first $1^+$ state of the intermediate nucleus $(A,Z+1)$ \cite{Abad1987}.
It can be extended to the more relaxed low-lying-states dominance (LLD) hypothesis \cite{Moreno2008}.
By analyzing the energy distribution of the emitted electrons,
the evidence of SSD for $^{82}$Se and $^{100}$Mo has been found
in CUPID \cite{Azzolini2019,Armengaud2020} and NEMO-3 \cite{Arnold2019}.
The mechanism of SSD (LLD) could be either
no contributions from highly lying $1^+$ states or the cancellations among them
\cite{Civitarese1998,Civitarese1999,Simkovic2001,Domin2005,Sarriguren2016,Simkovic2018}.
The running sum, namely, the cumulative contribution from the intermediate states,
of $M^{2\nu}_{\rm GT}$ is a useful tool in the study of SSD and LLD hypotheses \cite{Fang2010}.
It has been found that in the running sums,  calculated by quasiparticle random phase approximation (QRPA) \cite{Simkovic2018}
or nuclear shell model with a spin-orbit complete model space \cite{Horoi2007},
the contributions are positive at first
and then become negative from states lying around 6-10 MeV.
Thus, the appearance of the negative contributions seems to be a universal phenomenon \cite{Simkovic2018}.
And such negative contributions could lead to the cancellations among the higher-lying $1^{+}$ states
and consequently the fulfillment of SSD or LLD hypothesis. However, the underlying reason for the emergence of such negative contributions still needs investigations.

In QRPA model, the suppression of $M^{2\nu}_{\rm GT}$ is closely related to the isoscalar (IS) pairing,
as first identified by  Vogel and Zirnbauer in 1986 \cite{Vogel1986}.
By considering the particle-particle interaction,
mostly the IS pairing \cite{Rodin2011},
in the QRPA treatment of $2\nu\beta\beta$ decay,
the highly suppressed $M^{2\nu}_{\rm GT}$ of $^{130}$Te was successfully reproduced  \cite{Vogel1986}.
They also found that as a function of the strength of IS pairing,
the $M^{2\nu}_{\rm GT}$ of $^{130}$Te monotonously decreases and passes through zero,
which has been found later as a common feature in many nuclei in the following studies
\cite{Faessler1998,Muto1989,AlvarezRodriguez2004,Civitarese2000,Suhonen2005,Rodin2011,Simkovic2013}.
Such a suppression of $M^{2\nu}_{\rm GT}$ when increasing IS pairing strength
is found to be related with the restoration of the Wigner's spin-isospin SU(4) symmetry
\cite{Vogel1986,Rumyantsev1998,Civitarese2000,Rodin2005,Rodin2011}.
On the other hand, the ground-state correlations introduced by the backward amplitude
in QRPA calculations compared to that of quasiparticle-Tamm-Dancoff approximation (QTDA) model
was also shown to have the suppression effects on the NME \cite{Muto1989}.

The aim of this paper is to understand the mechanism of cancellation among higher-lying $1^{+}$ states that makes SSD and LLD hypotheses valid.
Inspired by the previous works on the suppression of  NME,
particular attentions will be paid on the roles of IS pairing as well as ground-state correlations on the negative contributions
in the running sums of $M^{2\nu}_{\rm GT}$.
During the past few decades, different nuclear models have been used in the study of $M^{2\nu}_{\rm GT}$
(see Refs. \cite{Faessler1998,Suhonen1998,Engel2017,Ejiri2019,Yao2021} for a review),
such as nuclear shell model \cite{Horoi2007,Caurier2012,Senkov2014,Brown2015,Li2017,Coraggio2019},
QRPA model
\cite{Vogel1986,Muto1989,AlvarezRodriguez2004,Civitarese2000,Suhonen2005,
Fang2010,Rodin2011,Suhonen2012,Sarriguren2012,Simkovic2013,Nicolas2015,Brown2015,Simkovic2018,Simkovic2018b,Terasaki2019,
Suhonen2014,Delion2015},
projected Hartree-Fock-Bogoliubov model \cite{Dixit2002,Chandra2005,Singh2007,Rath2010,Rath2019},
and interacting boson model \cite{Yoshida2013,Barea2015}.
One of the models which can avoid the closure approximation is the QRPA.
The closure approximation is not appropriate for $2\nu\beta\beta$ decay \cite{Suhonen1998}.
To study the running sum of NME, one must go beyond this approximation.

In this work, we adopt the spherical QRPA approach based on Skyrme Hartree-Fock-Bogoliubov model (Skyrme HFB + QRPA)
to systematically study the $M^{2\nu}_{\rm GT}$ of 11 nuclei with experimental NMEs.
We focus on the cancellation among higher-lying $1^{+}$ states in the running sum of $M^{2\nu}_{\rm GT}$ that leads to the fulfillment of SSD or LLD hypothesis.
Compared to QTDA model, the QRPA model introduces more ground-state correlations through the backward amplitude $Y_{\pi\nu}$.
We will study the importance of these ground-state correlations.
Their effects, entangled with the IS pairing, on the negative contributions
of the running sums for $M^{2\nu}_{\rm GT}$ are analyzed in detail.  The QRPA model as well as the formalism for calculating the $M^{2\nu}_{\rm GT}$ are presented in Sec.~\ref{secTheo}.
Numerical details of our Skyrme HFB+QPRA calculations are presented in Sec.~\ref{secNume}.
Sec.~\ref{secResu} contains the results and discussions.
The summary is given in Sec.~\ref{secConclu}.

\section{Formalism}
\label{secTheo}

\subsection{Quasiparticle random phase approximation}
\label{subsecQRPA}
We first perform the Skyrme HFB calculation, the formalism of which can be found in
Refs.~\cite{Dobaczewski1984,Dobaczewski1996,Bennaceur2005}.
Then with the obtained canonical single-particle wave functions and occupation amplitudes,
the QRPA equations for the state $|nJ^{P}\rangle$ with angular momentum $J$ and parity $P$
can be constructed in the angular momentum coupled form:
\begin{equation}
   \begin{aligned}
   \left(
     \begin{array}{cc}
        {A}        &  {B} \\
       -{B}        & -{A} \\
     \end{array}
   \right)
   \left(
     \begin{array}{c}
       {X}^{nJ^P} \\
       {Y}^{nJ^P} \\
     \end{array}
   \right)
   =&
   \Omega^{nJ^P}
   \left(
     \begin{array}{c}
       {X}^{nJ^P} \\
       {Y}^{nJ^P} \\
     \end{array}
   \right),
   \end{aligned}
   \label{Eq_QRPA}
\end{equation}
where $X^{nJ^P}$ and $Y^{nJ^P}$ are respectively the forward and backward amplitudes, and $\Omega^{nJ^P}$ are the energy eigenvalues.
In the charge-exchange case, using the canonical basis the matrices $A$ and $B$ can be expressed as
\begin{subequations}
\begin{equation}
   \begin{aligned}
        {A}_{\pi\nu,\pi'\nu'}
    =&
         \Big(
           u_{\pi} u_{\pi'} h_{\pi\pi'}
         - v_{\pi} v_{\pi'} h^{T}_{\bar{\pi} \bar{\pi}'} \\ &
         - u_{\pi} v_{\pi'} \Delta_{\pi \bar{\pi}'}
         + v_{\pi} u_{\pi'} \Delta^{\ast}_{\bar{\pi} \pi'}\Big) \delta_{\nu\nu'}  \\
     &+  \Big(
           u_{\nu} u_{\nu'} h_{\nu\nu'}
         - v_{\nu} v_{\nu'} h^{T}_{\bar{\nu} \bar{\nu}'} \\ &
         - u_{\nu} v_{\nu'} \Delta_{\nu \bar{\nu}'}
         + v_{\nu} u_{\nu'} \Delta^{\ast}_{\bar{\nu} \nu'} \Big) \delta_{\pi\pi'}\\
    &+    (u_{\pi} v_{\nu} u_{\pi'} v_{\nu'} + v_{\pi} u_{\nu} v_{\pi'}  u_{\nu'})
    \langle \pi \nu'| V |\nu \pi'\rangle^{ph}_J \\
    &+    (u_{\pi} u_{\nu} u_{\pi'} u_{\nu'} + v_{\pi} v_{\nu} v_{\pi'} v_{\nu'} )
    \langle \pi \nu | V |\pi' \nu'\rangle^{pp}_J      ,
   \end{aligned}
  \label{Eq_QRPA_A}
\end{equation}
\begin{equation}
   \begin{aligned}
    {B}_{\pi\nu,\pi'\nu'}  
       = & (v_{\pi} u_{\nu} u_{\pi'} v_{\nu'} + u_{\pi} v_{\nu} v_{\pi'} u_{\nu'})
         \langle \pi \nu'| V | \nu \pi'\rangle^{ph}_J \\
         &-(v_{\pi} v_{\nu} u_{\pi'} u_{\nu'} + u_{\pi} u_{\nu} v_{\pi'} v_{\nu'})
         \langle \pi \nu | V | \pi'\nu'\rangle^{pp}_J   ,
   \end{aligned}
  \label{Eq_QRPA_B}
\end{equation}
\end{subequations}
where $u$ and $v$ are the occupation amplitudes.
$h$ and $\Delta$ are respectively the single-particle Hamiltonian and pairing field.
$|\pi\rangle$ and $|\nu\rangle$ represent respectively the proton and neutron canonical states.
Their time reversal states are denoted as $|\bar{\pi}\rangle$ and $|\bar{\nu}\rangle$.
The residual interactions $V$
are divided into the particle-hole ($ph$) channel and the particle-particle ($pp$) channel respectively.
For the $ph$ channel, the same Skyrme interaction used in the HFB calculation is adopted.
For the $pp$ channel, namely the proton-neutron paring,
the density-dependent $\delta$ force is used,
which can be further divided into the isovector (IV) part
\begin{equation}
      V_{T=1}(\bm{r}_1, \bm{r}_2)
    =  f_{\rm IV}
     \left( t'_{0 \pi\nu}
      + \frac{t'_{3 \pi\nu}}{6}\rho(\bm{r})\right)
       \delta(\bm{r}_1 - \bm{r}_2)\frac{1-\hat{P}_{\sigma}}{2} ,
  \label{Eq_IVpairing}
\end{equation}
and the isoscalar (IS) part
\begin{equation}
      V_{T=0}(\bm{r}_1, \bm{r}_2)
    = f_{\rm IS} \left( t'_{0 \pi\nu}
     + \frac{t'_{3 \pi\nu}}{6}\rho(\bm{r})\right)
      \delta(\bm{r}_1 - \bm{r}_2) \frac{1+\hat{P}_{\sigma}}{2} ,
  \label{Eq_ISpairing}
\end{equation}
where $\bm{r} = (\bm{r}_1+\bm{r}_2)/2$, and $\hat{P}_{\sigma}$ is the spin-exchange operator.
Parameters $t'_{0 \pi\nu}$ and $t'_{3 \pi\nu}$
are set as the average of the proton-proton and neutron-neutron pairing strengths, which are
determined from the Skyrme HFB calculations to reproduce the proton and neutron pairing gaps
from the five-point formula of experimental binding energies respectively.
Then in QRPA calculations, $f_{\rm IV}$ is fixed to make the Fermi $2\nu\beta\beta$ matrix element vanish
as it should, due to the different isospin quantum numbers of the mother nucleus ($T$) and daughter nucleus ($T-2$).
The strength of isoscalar proton-neutron paring $f_{\rm IS}$ is not well constrained \cite{Sagawa2016},
it is usually fixed through fitting the experimental NME.
We note that the isovector proton-neutron pairing $V_{T=1}$ does not affect
the Gamow-Teller $2\nu\beta\beta$ matrix element $M^{2\nu}_{\rm GT}$ \cite{Simkovic2013}.

\subsection{$2\nu\beta\beta$-decay nuclear matrix element}
\label{subsecNME}
For the $2\nu\beta\beta$ decay, due to isospin symmetry, the NME is dominated by GT transitions, $M^{2\nu}_{\rm GT}$.
For the ground-state-to-ground-state $2\nu\beta\beta$ decay, $M^{2\nu}_{\rm GT}$ can be expressed as
\begin{equation}
  \begin{aligned}
   M_{\rm GT}^{2\nu}
    =&  \sum_{n_i n_f}
             \frac{ \langle 0^{+(f)}_{\rm g.s.} || \hat{O}^{-}_{\rm GT} || 1^{+}_{n_f} \rangle
                     \langle 1^{+}_{n_f} | 1^{+}_{n_i} \rangle
                     \langle 1^{+}_{n_i} || \hat{O}^{-}_{\rm GT} || 0^{+(i)}_{\rm g.s.} \rangle }
                    {E_{\rm int.}(n_i, n_f) + M_{\rm int.} - (M_{f}+M_{i})/2} ,
  \end{aligned}
  \label{Eq_NME}
\end{equation}
where $\langle 1^{+}_{n_i} || \hat{O}^{-}_{\rm GT} || 0^{+(i)}_{\rm g.s.} \rangle$
is the GT$^{-}$ transition amplitude of the initial nucleus to the intermediate nucleus. It can be obtained from QRPA model as
\begin{equation}
 \begin{aligned}
  \langle 1^{+}_{n_i} || \hat{O}^{-}_{\rm GT} || 0^{+(i)}_{\rm g.s.} \rangle
=& \sum_{\pi_i \nu_i} - \langle j_{\pi_i} ||\hat{O}^{-}_{\rm GT}|| j_{\nu_i} \rangle \\ & \times
       \left( X^{n_i}_{\pi_i \nu_i} u_{\pi_i} v_{\nu_i} + Y^{n_i}_{\pi_i \nu_i}  v_{\pi_i} u_{\nu_i} \right).
  \label{Eq_GT-_i}
 \end{aligned}
\end{equation}
And the transtion from the intermediate states to the final nucleus $\langle 0^{+(f)}_{\rm g.s.} || \hat{O}^{-}_{\rm GT} || 1^{+}_{n_f} \rangle$ can be expressed as
\begin{equation}
 \begin{aligned}
   \langle 0^{+(f)}_{\rm g.s.} || \hat{O}^{-}_{\rm GT} || 1^{+}_{n_f} \rangle
=& -\langle 1^{+}_{n_f} || \hat{O}^{+}_{\rm GT} || 0^{+(f)}_{\rm g.s.} \rangle    \\
=& \sum_{\pi_f \nu_f}-\langle j_{\pi_f} ||\hat{O}^{+}_{\rm GT}|| j_{\nu_f} \rangle \\ & \times
   \left( X^{n_f}_{\pi_f \nu_f} v_{\pi_f} u_{\nu_f} + Y^{n_f}_{\pi_f \nu_f} u_{\pi_f} v_{\nu_f} \right).
 \end{aligned}
  \label{Eq_GT+_f}
\end{equation}
The overlap factor between the intermediate states constructed from the initial and final nuclei $\langle 1^{+}_{n_f} | 1^{+}_{n_i} \rangle$ is \cite{Simkovic2004},
\begin{equation}
   \begin{aligned}
      \langle 1^{+}_{n_f}|1^{+}_{n_i} \rangle
   =&  \sum_{\pi_i \nu_i}  \sum_{\pi_f \nu_f} C_{\pi_i \pi_f} C_{\nu_i \nu_f}
    (X^{n_i}_{\pi_i \nu_i} X_{\pi_f \nu_f}^{n_f} - Y^{n_i}_{\pi_i \nu_i} Y_{\pi_f \nu_f}^{n_f}) \\
    &~~~~~~~~ \times
        (u_{\pi_i} u_{\pi_f} + v_{\pi_i} v_{\pi_f})
        (u_{\nu_i} u_{\nu_f} + v_{\nu_i} v_{\nu_f})  \\
    &~~~~~~~~ \times
        \langle {\rm{HFB}}^{(f)}|{\rm{HFB}}^{(i)} \rangle ,
   \end{aligned}
\end{equation}
where $C_{\pi_i \pi_f} = \langle \pi_i | \pi_f \rangle $
and $C_{\nu_i \nu_f} = \langle \nu_i | \nu_f \rangle $
are the overlaps of the canonical single-particle wave functions.
In the canonical basis, the overlap of the HFB ground states of initial and final nuclei reads
\begin{equation}
  \begin{aligned}
       \langle {\rm{HFB}}^{(f)}|{\rm{HFB}}^{(i)} \rangle
    =  \prod_{ \pi_i \pi_f \nu_i \nu_f}
     &     (u_{\pi_i} u_{\pi_f} + v_{\pi_i} v_{\pi_f})  \\ \times
     &     (u_{\nu_i} u_{\nu_f} + v_{\nu_i} v_{\nu_f}).
  \end{aligned}
\end{equation}
The excitation energy of intermediate nucleus in the denominator is
\begin{equation}
	\label{intE}
   \begin{aligned}
   E_{\rm int.}(n_i, n_f)
    =& \frac{1}{2} \left(E_{\rm int.}^{n_i} + E_{\rm int.}^{n_f} \right)                  \\
    =& \frac{1}{2}\Big[     (\Omega^{n_i} + \lambda^{(i)}_{\pi} - \lambda^{(i)}_{\nu} )
                           - (M_{\rm int.} - M_i) - \Delta m_{\nu\pi}   \\
     &                     +(\Omega^{n_f} - \lambda^{(f)}_{\pi} + \lambda^{(f)}_{\nu} )
                           - (M_{\rm int.} - M_f) + \Delta m_{\nu\pi} \Big]  \\
                           =& \frac{1}{2}\Big[     (\Omega^{n_i} + \lambda^{(i)}_{\pi} - \lambda^{(i)}_{\nu} )
                           +(\Omega^{n_f} - \lambda^{(f)}_{\pi} + \lambda^{(f)}_{\nu} )  \\
     &                     - (2M_{\rm int.} - M_i - M_f)  \Big]  ,
   \end{aligned}
\end{equation}
where $\lambda$ denotes the Fermi surface and $\Delta m_{\nu\pi}$
is the mass difference between the neutron and the proton, i.e., $m_{\nu} - m_{\pi}$.
$\Omega^{n_i}$ and $\Omega^{n_f}$ are the eigenvalues of QRPA equations for mother and daughter nuclei, respectively.
With the use of Eq. \eqref{intE}, nuclear masses are eventually not needed in the calculation of NME in Eq. \eqref{Eq_NME},
only the energy with respect to the mother nucleus $(\Omega^{n_i} + \lambda^{(i)}_{\pi} - \lambda^{(i)}_{\nu} )$
and the energy with respect to the daughter nucleus $(\Omega^{n_f} - \lambda^{(f)}_{\pi} + \lambda^{(f)}_{\nu} )$ are needed,
which are obtained from QRPA calculations.
However, to calculate the energy of intermediate nucleus $E_{\rm int.}(n_i, n_f)$
the experimental mass values from AME2020 \cite{Huang2021} are used
for the masses of the initial nucleus $M_i$, the intermediate nucleus $M_{\rm int.}$, and the final nucleus $M_f$.

\section{Numerical details}
\label{secNume}
For the Skyrme HFB calculations,
the SkO$'$ interaction \cite{Reinhard1999} is used for the mean-field calculation.
For the pairing interaction, we adopt the surface density-dependent $\delta$ force, whose strength is fixed to reproduce
the experimental pairing gap obtained from the five-point formula of binding energies as stated above.
We use different cut-offs to reduce the computation burden, and obtain required accuracy.
For the mean field, we use a smooth cut-off for the pairing window with a diffuseness parameter $\mu$ being 0.1 MeV,
and the cut-off on equivalent Hartree-Fock energy $\varepsilon^{\rm{HF}}$ is set to be 80.0 MeV.
For the QRPA calculation, we use the canonical basis  ($|\pi\rangle$ and $|\nu\rangle$)
with occupation amplitudes ($u$ and $v$), which are obtained by performing canonical transformation
on the quasiparticle states in HFB calculations.
The $\pi$-$\nu$ configurations are selected under the criteria $|u_{\pi} v_{\nu}|>10^{-4}$
or $|u_{\nu} v_{\pi}|>10^{-4}$, with the single-particle energies in canonical basis $\varepsilon_{\pi(\nu)}<60.0$ MeV.
For the calculations of $M^{2\nu}_{\rm GT}$ of specific intermediate states,
only the $\pi$-$\nu$ configurations with $|X^2_{\pi\nu}-Y^2_{\pi\nu}|>10^{-6}$ are taken.
The strengths of isovector proton-neutron paring $f_{\rm IV}$ are fixed by
tuning $M_{\rm F}^{2\nu}$ vanish.
But since it does not affect
the Gamow-Teller $2\nu\beta\beta$ matrix element $M^{2\nu}_{\rm GT}$,
we will not discuss it in present work.

\section{Results and discussions}
\label{secResu}

\subsection{Systematic study on  $M_{\rm GT}^{2\nu}$}
\label{secSys}
\begin{figure}[t]
  \centering
  \includegraphics[width=0.45\textwidth]{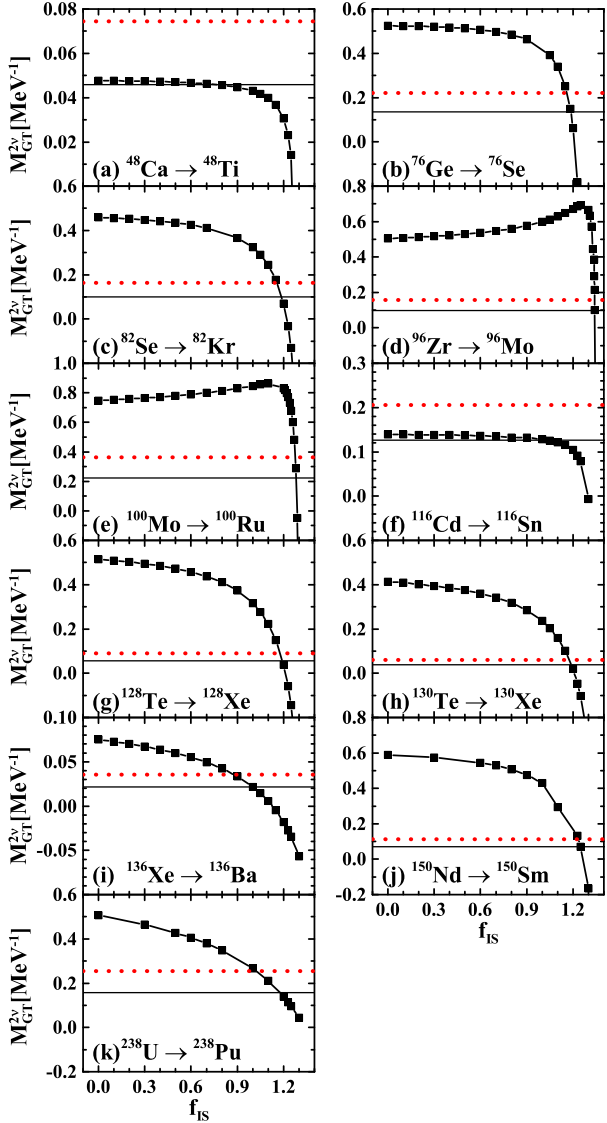}\\
  \caption{The $2\nu\beta\beta$ nuclear matrix elements $M_{\rm GT}^{2\nu}$ for 11 nuclei
           as a function of isocalar pairing strength parameter (black square).
           The horizontal lines represent the experimental values extracted from Ref. \cite{Barabash2015}
           with $g_A = 1.00$ (red dotted), and $g_A=1.27$ (black dashed)}.
           \label{Fig1}
\end{figure}

We start with systematic study of the evolution of $2\nu\beta\beta$-decay
NMEs ($M_{\rm GT}^{2\nu}$) on the IS pairing strength for 11 nuclei which have been measured.
Results are depicted in Fig.~\ref{Fig1} and compared with compiled data from
Ref. \cite{Barabash2015}.
From the figure, all these $M_{\rm GT}^{2\nu}$ decrease when IS pairing strength is large enough.
This suppression effect of IS pairing, as a common feature found in QRPA calculations,
has been well analyzed in Refs.~\cite{Vogel1986,Civitarese1987}.
The decreasing behavior of the $M_{\rm GT}^{2\nu}$ with increasing IS paring strength
enables us to find the appropriate strength parameter $f_{\rm IS}$
to reproduce the experimental values, as listed in Tab.~\ref{Tab_1}.
If one further increases the IS pairing strength, the $M^{2\nu}_{\rm GT}$ passes through zero and changes its sign.
It has been found that the broken Wigner spin-isospin SU(4) symmetry is restored
when the $2\nu\beta\beta$ closure matrix element
$M_{{\rm GT} cl}^{2\nu}= \langle f| \sum_{mn} \vec{\sigma}_m \cdot \vec{\sigma}_{n} \tau_m^{+} \tau_{n}^{+} |i\rangle=0$
\cite{Rumyantsev1998,Civitarese2000,Rodin2005,Rodin2011}.
We note that besides the IS pairing,
the nuclear deformation also leads to the suppression of the $M^{2\nu}_{\rm GT}$ \cite{Simkovic2004,AlvarezRodriguez2004},
mostly due to the difference between the shapes of initial and final nuclei,
which will not be discussed in present work.

In Tab.~\ref{Tab_1}, we list the experimental and theoretical values of $M^{2\nu}_{\rm GT}$ for these 11 nuclei.
The isoscalar pairing strength parameters $f_{\rm IS}$ used in QRPA calculations are listed in the last column.
We observe that most of the determined $f_{\rm IS}$ are around 1.2, which means the ground-state correlation
should be large enough to suppress the $M^{2\nu}_{\rm GT}$.
For the nuclei with magic numbers, $^{48}$Ca, $^{116}$Cd, and $^{136}$Xe, their $f_{\rm IS}$ values are relatively small.
This is probably caused by the overestimated small ground-state overlap factor
due to the violation of particle number in QRPA model
\cite{Yao2015} and this needs further investigation.
We further examine the SSD and the extended LLD hypotheses, so
the $M^{2\nu}_{\rm GT}$ obtained from the SSD or LLD hypothesis
$M_{\rm GT}^{2\nu}({\textrm{SSD or LLD}})$ are also listed in Tab~\ref{Tab_1}.
When $g_A=1.27$ (bare value), for $^{48}$Ca, $^{82}$Se, $^{116}$Cd, $^{128}$Te, and $^{238}$U,
the first $1^+$ states of the intermediate nuclei contribute more than $75\%$ to the total NMEs.
In this sense, the SSD hypothesis is fulfilled for these nuclei.
On the other hand, when $g_A=1.00$ (quenched value), only $^{238}$U still fulfills the SSD hypothesis.
The inconsistence on the evidence of SSD in $^{100}$Mo reported in Refs. \cite{Arnold2019,Armengaud2020}
could be the nuclear deformation that is not included in our model \cite{Moreno2008}.
We further test the LLD hypothesis for the nuclei whose NMEs do not fulfill the SSD hypothesis.
Because there is not a definite upper limit of the energy for low-lying states in intermediate nuclei,
we set it as 5 MeV.
When $g_A=1.00$, these low-lying states contribute almost the full NMEs for $^{76}$Ge, and $^{82}$Se.
In this sense, the NMEs for these nuclei fulfill the LLD hypothesis.

\begin{table}[t]
    \caption{
    Experimental and theoretical $2\nu\beta\beta$ nuclear matrix elements $M_{\rm GT}^{2\nu}$ for 11 nuclei with unit of MeV$^{-1}$
    when $g_A=1.27$ and $g_A=1.00$.
    $M_{\rm GT}^{2\nu}$ obtained by the single-state dominance (SSD)
    and low-lying-states dominance (LLD) hypotheses are also listed.
    The upper limit of the energy for low-lying states $E_{\rm int.}$ in LLD is set as 5 MeV.
    The isocalar pairing strength parameters used in QRPA calculations are listed in the last column.
    The experimental values of $M_{\rm GT}^{2\nu}$ are taken from Ref.\cite{Barabash2015}.
    The results of $^{48}$Ca and $^{116}$Cd when $g_A=1.00$ are not listed,
    because the experimental NMEs are always larger than the theoretical ones with this value of $g_A$.}
    \centering
  \resizebox{0.48\textwidth}{!}{
  \begin{tabular}{ccccccc}
  \hline\hline
            &              &  Expt.   &   Theo.    &    Theo.    &   Theo.   &   \\
   Nucleus                 &
  $g_A$      &
  $M_{\rm GT}^{2\nu}$      &
  $M_{\rm GT}^{2\nu}$      &
  $M_{\rm GT}^{2\nu}$(SSD) &
  $M_{\rm GT}^{2\nu}$(LLD) &
   $f_{\rm IS}$            \\ \hline
  $^{48}$Ca & 1.27 &  0.046$\pm$0.004  &   0.046    &    0.036    &   0.036   &  0.7570  \\
  $^{76}$Ge & 1.27 &  0.137$\pm$0.007  &   0.136    &    0.071    &   0.188   &  1.1852  \\
            & 1.00 &  0.221$\pm$0.012  &   0.221    &    0.082    &   0.256   &  1.1612  \\
  $^{82}$Se & 1.27 &  0.101$\pm$0.005  &   0.100    &    0.104    &   0.161   &  1.1886  \\
            & 1.00 &  0.162$\pm$0.008  &   0.162    &    0.095    &   0.191   &  1.1594  \\
  $^{96}$Zr & 1.27 &  0.097$\pm$0.005  &   0.099    &    0.351    &   0.252   &  1.3411  \\
            & 1.00 &  0.157$\pm$0.008  &   0.158    &    0.405    &   0.309   &  1.3399  \\
  $^{100}$Mo& 1.27 &  0.224$\pm$0.006  &   0.224    &    0.515    &   0.346   &  1.2824  \\
            & 1.00 &  0.362$\pm$0.010  &   0.361    &    0.631    &   0.478   &  1.2768  \\
  $^{116}$Cd& 1.27 &  0.127$\pm$0.004  &   0.127    &    0.112    &   0.124   &  1.0350  \\
  $^{128}$Te& 1.27 &  0.056$\pm$0.007  &   0.057    &    0.043    &   0.113   &  1.1928  \\
            & 1.00 &  0.090$\pm$0.012  &   0.091    &    0.047    &   0.130   &  1.1788  \\
  $^{130}$Te& 1.27 &  0.038$\pm$0.005  &   0.037    &    0.020    &   0.087   &  1.1912  \\
            & 1.00 &  0.061$\pm$0.008  &   0.062    &    0.022    &   0.095   &  1.1776  \\
  $^{136}$Xe& 1.27 &  0.022$\pm$0.001  &   0.022    &    0.003    &   0.007   &  0.9968  \\
            & 1.00 &  0.035$\pm$0.001  &   0.035    &    0.004    &   0.011   &  0.8820  \\
  $^{150}$Nd& 1.27 &  0.070$\pm$0.005  &   0.070    &    0.240    &   0.257   &  1.2503  \\
            & 1.00 &  0.114$\pm$0.008  &   0.116    &    0.246    &   0.289   &  1.2360  \\
  $^{238}$U & 1.27 &  0.158$^{+0.109}_{-0.085}$  &   0.157    &    0.185    &   0.185   &  1.1787  \\
            & 1.00 &  0.254$^{+0.176}_{-0.137}$  &   0.254    &    0.230    &   0.230   &  1.0251  \\
  \hline \hline
\end{tabular}}
\label{Tab_1}
\end{table}

\begin{figure}[!t]
  \centering
  \includegraphics[width=0.45\textwidth]{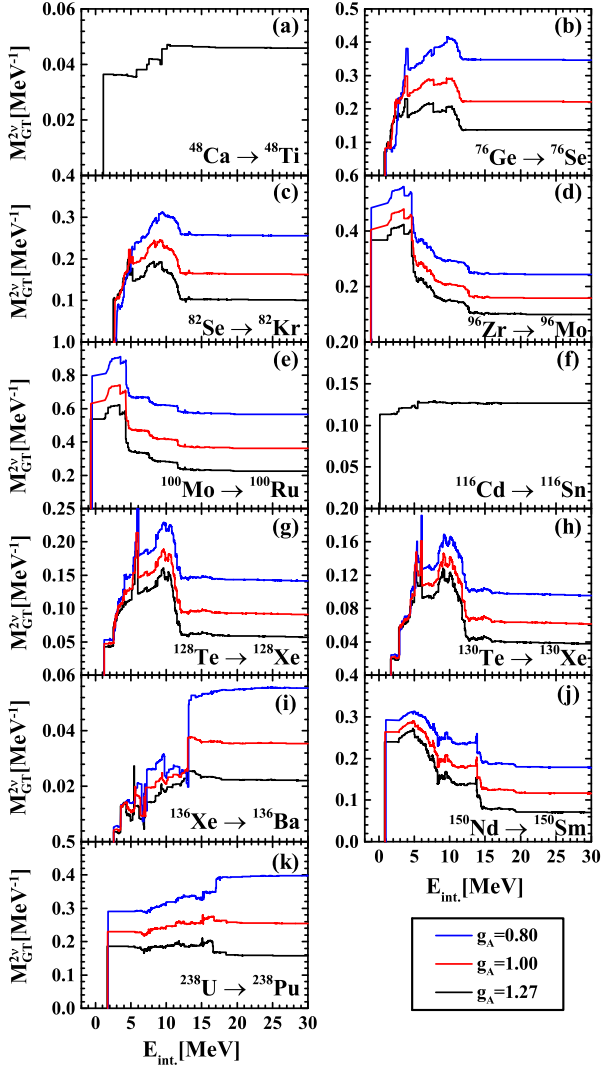}\\
  \caption{
  Running sums of $M_{\rm GT}^{2\nu}$ for 11 nuclei
  as a function of the excitation energy of the intermediate nucleus
  with $g_A = 0.80$ (blue), $g_A = 1.00$ (red), and $g_A=1.27$ (black).
  The IS pairing strength parameters $f_{\rm IS}$ used for the corresponding calculations are also shown in the figure. }
  \label{Fig2}
\end{figure}

In Fig.~\ref{Fig2}, with the determined $f_{\rm IS}$ in Tab.~\ref{Tab_1},
the running sums of  $M^{2\nu}_{\rm GT}$ for these 11 nuclei are depicted.
The results of $g_A=0.80$ are also shown, since this value of $g_A$ has been used in Ref. \cite{Gando2019}.
Generally, the behaviors of running sums can be divided into three types when $g_A=1.27$.
The first one is steadily increasing accompanied with small fluctuations,
as shown in $^{48}$Ca and $^{116}$Cd.
The second type is that the contributions from first few lowest states are large enough
but then the following excited states give continuous negative contributions, and
hence provide a suppression due to the cancellation between lowest states and higher-lying states.
like $^{96}$Zr, $^{100}$Mo, or $^{150}$Nd.
The last type is the most common one, observed in 6 of  total 11 nuclei, i.e.,
$^{76}$Ge, $^{82}$Se, $^{128}$Te, $^{130}$Te, $^{136}$Xe, and $^{238}$U.
The cumulative contributions continuously increase up to
the excitation energy of the intermediate nucleus $E_{\rm int.}$ around 10 MeV (for $^{136}$Xe and $^{238}$U, it is 15 MeV),
then a remarkable cancellation appears due to the negative contributions of higher-lying states,
being similar as the case of the second type.
The difference between the second and the third type is how the medium energy states with $5$-$10$ MeV contribute.
The cancellation between positive contributions from the low-lying states
and negative contributions from the higher-lying states
leads to the realization of SSD for $^{82}$Se and $^{128}$Te when $g_A=1.27$, and
LLD for $^{76}$Ge and $^{82}$Se when $g_A=1.00$.
The negative contributions in the running sums seem to be universal,
which also appear in the results calculated by shell model
with complete spin-orbit partner model space \cite{Nakada1996,Horoi2007}
and QRPA models \cite{Fang2010,Suhonen2014,Delion2015,Simkovic2018}.
Compared to the nuclei of the second and third types, it is noticed that
the IS pairing strength parameters $f_{\rm IS}$ are smaller
for the nuclei of the first type.
On the other hand,
from Fig. \ref{Fig1}, when increasing $g_A$, the strength of proton-neutron isoscalar pairing
$f_{\rm IS}$ should also be increased to reproduce the experimental NME.
By comparing the running sums of $g_A=0.80, 1.00$ and $1.27$, one can find that
when $g_A$ is increasing
the negative contributions can either be induced, e.g. in $^{136}$Xe and $^{238}$U,
or be enlarged, e.g. in $^{82}$Se, $^{128}$Te, and $^{130}$Te.
Therefore, these two aspects indicate that the appearance of negative contributions in the running sums
are closely related with the magnitude of $f_{\rm IS}$, which controls
the amount of ground-state correlations introduced in QRPA \cite{Fang2010}.
So, without losing generality, we will pick up $^{128}$Te as an example and analyze the mechanism for the negative contributions of high-lying states at large $f_{\rm IS}$ in next subsections.

\subsection{Ground-state correlations: QRPA vs. QTDA}
\label{secGSC}
\begin{figure}[!t]
  \centering
  \includegraphics[width=0.4\textwidth]{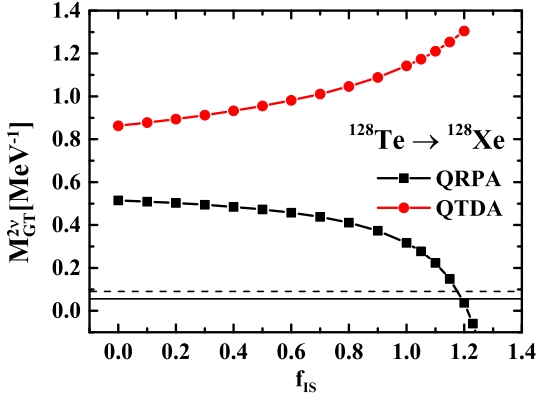}\\
  \caption{The $2\nu\beta\beta$ nuclear matrix elements $M_{\rm GT}^{2\nu}$ for $^{128}$Te
                 as a function of isocalar pairing strength parameter calculated by QRPA (black square)
                 and QTDA (red circle).
                 The horizontal lines represent the experimental values with $g_A=1.27$ (solid) and $g_A=1.00$ (dashed).}
                 \label{Fig3}
\end{figure}

As stated above, the negative contributions in the running sums
are related with the ground-state correlations in the QRPA model.
In this subsection, to analyze the effects of the ground-state correlations,
as an example, we compare the QRPA model and the QTDA (quasiprticle Tamm-Dancoff approximation) model calculations of the $M^{2\nu}_{\rm GT}$ for $^{128}$Te.
Unlike the QRPA model, the ground state of the QTDA model is the quasiparticle vacuum with no ground-state correlation,
that is without the backward amplitude $Y_{\pi\nu}$ term.
In Fig.~\ref{Fig3}, the evolution of $M^{2\nu}_{\rm GT}$ as a function of
IS pairing strength $f_{\rm IS}$ from QRPA and QTDA  models are shown.
These two results are different from each other.
The $M^{2\nu}_{\rm GT}$ calculated by QTDA model monotonously increases with increasing $f_{\rm IS}$
and is systematically larger than that from QRPA model.
The latter shows explicitly the suppression effect from the ground-state correlations.
This is consistent with the conclusion of Ref.~\cite{Muto1989}.

\begin{figure}[!t]
  \centering
  \includegraphics[width=0.4\textwidth]{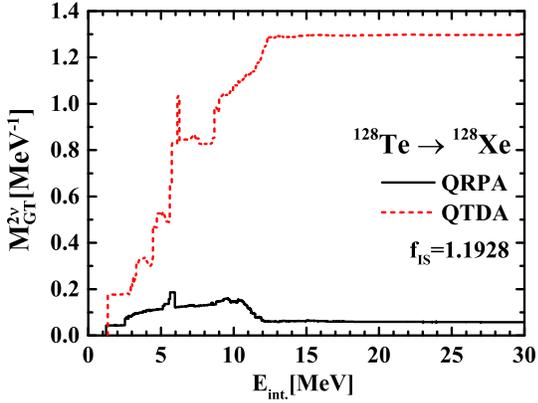}\\
  \caption{  Running sums of the $M_{\rm GT}^{2\nu}$ for $^{128}$Te
             as a function of the excitation energy of the intermediate nucleus
             calculated by QRPA (black solid line) and QTDA (red dashed line)
             with the isoscalar pairing strength parameter $f_{\rm IS}=1.1928$.}
             \label{Fig4}
\end{figure}

In Fig.~\ref{Fig4}, the running sums of $M^{2\nu}_{\rm GT}$ calculated by QTDA model and QRPA model with $f_{\rm IS}=1.1928$ which reproduce the experimental results for QRPA calcualtion, are shown respectively.
One finds that in the region of $E_{\rm int.} < 9.0$ MeV,
the running sums of these two models are different in magnitude but evolve similarly.
However, in the region of $9.0 < E_{\rm int.} < 12.0$ MeV,
the running sum of the QTDA model keeps increasing, while that of the QRPA model starts to decrease.
Therefore, the negative contributions of high-lying states are related with the ground-state correlations introduced in QRPA model.
Finally, running sum curves of both models become flat
after $E_{\rm int.}$ equals to around 12.0 MeV around the GT resonance (GTR) region,
because of large energy denominators and small transition amplitudes for these very highly excited states beyond GTR.

\begin{figure}[!t]
  \centering
  \includegraphics[width=0.4\textwidth]{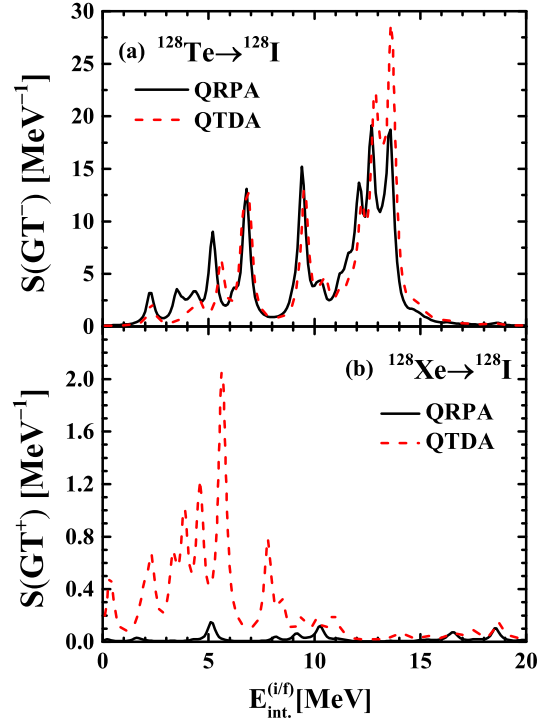}\\
  \caption{The GT$^{-}$ transition strength function of $^{128}$Te $\rightarrow$ $^{128}$I
           whose excitation energies $E_{\rm int.}^{(i)}$ are respect to $^{128}$I (a)
           and GT$^{+}$ transition strength function of $^{128}$Xe $\rightarrow$ $^{128}$I
           whose excitation energies $E_{\rm int.}^{(f)}$ are respect to $^{128}$I(b).
           The black solid lines and red dashed lines are
           the results calculated by QRPA and QTDA, respectively.}
           \label{Fig5}
\end{figure}

To understand the large difference between the $M^{2\nu}_{\rm GT}$ calculated by QRPA and QTDA models,
we further investigate GT$^{-}$ and GT$^{+}$ transition branches
involved in the calculation of  $M^{2\nu}_{\rm GT}$ as shown in Eq.~(\ref{Eq_NME}).
The corresponding transition strengths $S({\rm GT}^{-})$ and $S({\rm GT}^{+})$ are shown in Fig.~\ref{Fig5}.
Since $^{128}$Te is with neutron excess, its GT$^-$ transition strength is large, so the inclusion of $Y_{\pi\nu}$ amplitude
in QRPA calculation does not make it different from QTDA calculation without $Y_{\pi\nu}$ term.
However, the GT$^+$ transition strength of $^{128}$Xe is small,
due to the blocking effects of Pauli principle with large neutron excess.
The inclusion of
ground-state isovector pairing correlations
could unblock some transition channels and increase the GT$^+$ transition strength.
The inclusion of the backward amplitude $Y_{\pi\nu}$,
the ground-state correlations introduced in QRPA model,
have also significant influence on the GT$^+$ transition strength through the relatively big $u_{\pi}v_{\nu}$ factor
in front of $Y_{\pi\nu}$.
Due to the opposite signs of $X_{\pi\nu}$ and $Y_{\pi\nu}$,
for $^{128}$Xe the GT$^+$ transition strength becomes smaller in QRPA calculation compared to that of QTDA calculation.
Through the above comparison,
it is clear the ground-state correlations,
namely the large backward amplitudes $Y_{\pi\nu}$ in QRPA induced by strong IS pairing,
play crucial roles in the calculation of $M^{2\nu}_{\rm GT}$
through their effects on the GT$^{+}$ transition of daughter nucleus.

\subsection{Negative contributions in running sum}
\label{secNeg}
\begin{figure}[!t]
  \centering
  \includegraphics[width=0.4\textwidth]{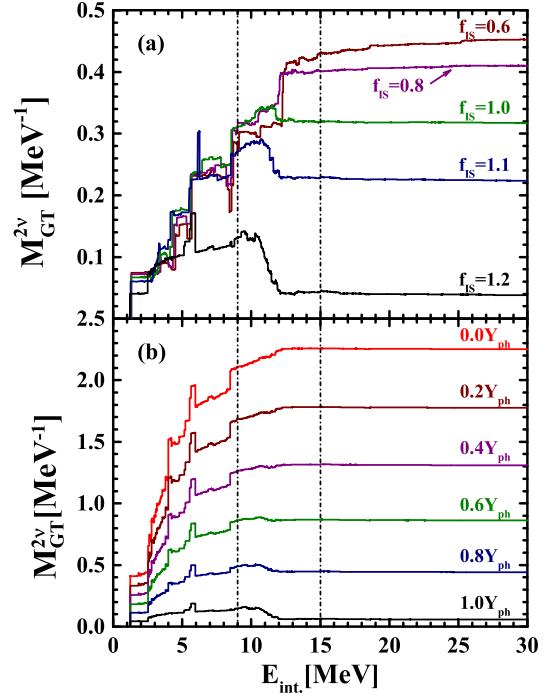}\\
  \caption{Running sums of the $M_{\rm GT}^{2\nu}$ for $^{128}$Te
           with different $f_{\rm IS}$ (a) and $\eta_{Y}$ (b)
           as a function of the excitation energy of the intermediate nucleus.}
           \label{Fig6}
\end{figure}

By comparing running sums of QRPA model and QTDA model in Sec.~\ref{secGSC}, we come to the conclusion that
the negative contributions are related with ground-state correlations
introduced by $Y_{\pi \nu}$ term which can be enhanced by large IS pairing.
Actually, there is a close relation between IS pairing and ground-state correlations introduced by $Y_{\pi \nu}$ term.
With the increase of IS pairing strength, the magnitude of $B$ matrix elements in QRPA equation \eqref{Eq_QRPA_B} increases as well
due to the attractive nature of IS pairing residual interaction.
As a result, the $Y_{\pi \nu}$  amplitude will increase, as also explained in Ref. \cite{Vogel1986}.
We depict the running sums of the $M^{2\nu}_{\rm GT}$ for $^{128}$Te with different
$f_{\rm IS}$ and $\eta_{Y}$ in Fig.~\ref{Fig6}.
From the figure, the behaviors of the running sum curves are very similar when increasing $f_{\rm IS}$ and $\eta_{Y}$.
Especially, in both cases the negative contributions will be induced and further be enlarged in the 9-15 MeV (around the GTR) region.
Based on such a relation,
to simulate the increase of the IS pairing strength,
we change the magnitude of $Y_{\pi\nu}$ by replacing it with $\eta_{Y} Y_{\pi\nu}$ in the calculation of transition amplitudes,
where the factor $\eta_Y$ varies from 0.0 to 1.0.
The purpose of this study is to understand how negative contributions to  $M^{2\nu}_{\rm GT} $ happen at large IS pairing strength.
One of the advantages of such a simulation, instead of varying the IS pairing strength directly,  is that one can avoid
the complicated splittings and degeneracy in the excited spectra caused by the variation of the IS pairing.

\begin{figure}[!t]
  \centering
  \includegraphics[width=0.4\textwidth]{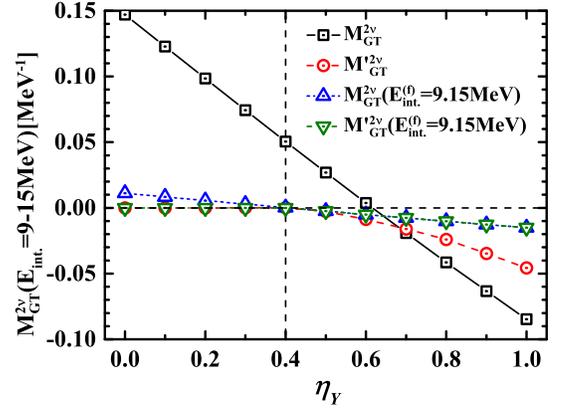}\\
  \caption{Contributions to $M_{\rm GT}^{2\nu}$ for $^{128}$Te
           from excitation energies of the intermediate nucleus ranging 9.0-15.0 MeV.
           Black square: the total contribution of  all intermediate states in the
           9.0-15.0 MeV excitation energy region.
           Red circle: sum of the contributions  whose signs   have changed relative to $\eta_Y=0.0$ during $\eta_Y$ increasing.
           Blue up triangle: sum of the contributions of those intermediate states with one coming from GT$^+$ transition of the daughter nucleus with an excitation energy of $E_{\rm int.}^{(f)}=9.15$ MeV.
           Green down triangle: sum of the contributions with
           $E_{\rm int.}^{(f)}=9.15$ MeV whose signs have changed relative to $\eta_Y=0.0$ during $\eta_Y$ increasing.}
           \label{Fig7}
\end{figure}

The trend of the results with $\eta_{Y}=0.0$ is very similar to that of QTDA model in Fig.~\ref{Fig4}.
With the increasing $\eta_{Y}$, two features of the running sum curves are emerging.
The first one is that the contributions from the region
$E_{\rm int.} < 9.0$ MeV reduce significantly.
The second one is that the majority of these contributions in the region of
$9.0 < E_{\rm int.} < 15.0$ MeV
changes their signs from positive to negative values. It can be seen that negative contributions do appear with large enough $Y_{\pi\nu}$ amplitudes in our simulation.
Then what is the origin of these negative contributions? To answer this question, in Fig.~\ref{Fig7},
we explicitly show the contributions to $M^{2\nu}_{\rm GT}$ of intermediate states
with excitation energy ranging from 9.0-15.0 MeV as a function of $\eta_Y$.
The total contribution (black square) decrease monotonously  from positive to negative values.
Then we separate the contributions which have changes in sign for the increasing $\eta_Y$,
denoted as ${M'}^{ 2\nu }_{\rm GT}$ (red circle).
From the figure, we observe that the ${M'}^{ 2\nu }_{\rm GT}$ contributes remarkably
more than 50\% to the ${M}^{ 2\nu }_{\rm GT}$  at $\eta_Y = 1.0$. So a main reason that the negative contributions appear is
the change of signs of $M^{2\nu}_{\rm GT}$ contributed by some intermediate states.
Among these intermediate states, one non-negligible contribution comes from those with an excitation energy of
$E_{\rm int.}^{(f)}=9.15$ MeV from the GT$^+$ transition side of the daughter nucleus.
We note that, the rest part of the negative contributions
do not change their signs for the increasing $\eta_Y$,
but they are small at $\eta_Y=0.0$ and become large at $\eta_Y=1.0$.

\begin{figure}[t]
  \centering
  \includegraphics[width=0.4\textwidth]{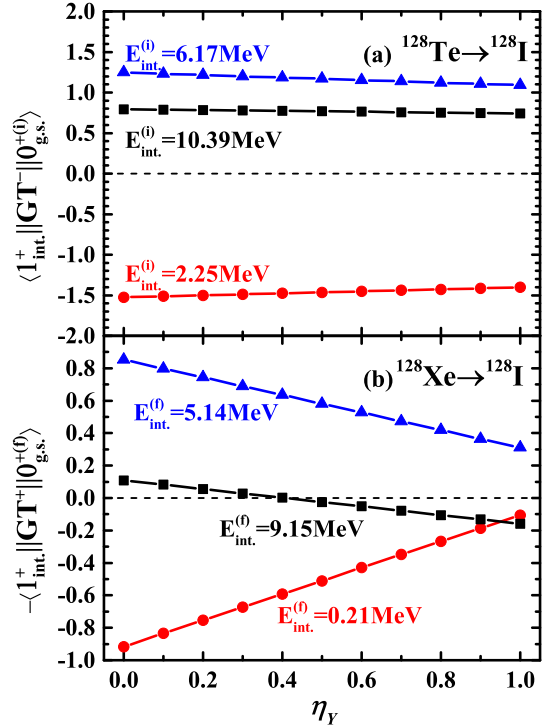}\\
  \caption{The GT$^{-}$ transition amplitudes for 3 excited states in $^{128}$Te (a)
           and GT$^{+}$ transition amplitudes for 3 excited states in $^{128}$Xe (b)
           as a function of $\eta_Y$.
           The values of excitation energies relative to $^{128}$I are shown in the figure.}
           \label{Fig8}
\end{figure}

To get a deeper insight for these sign changes,
we investigate the GT$^{-}$ and GT$^{+}$ transitions, respectively.
From Eq.~(\ref{Eq_NME}), the ${M}^{ 2\nu }_{\rm GT}$ for a specific intermediate state is proportional to
the GT$^-$ transition amplitude $\langle 1^{+}_{\rm int.}|| {\rm GT}^{-} || 0^{+(i)}_{\rm g.s.}\rangle$
and the GT$^+$ transition amplitude $\langle 1^{+}_{\rm int.}|| {\rm GT}^{+} || 0^{+(f)}_{\rm g.s.}\rangle$.
So we pick up several typical GT$^-$ and GT$^+$ states with excitation energies in the range of $E_{\rm int.}  < 9.0$ MeV
and in the range of $9.0 < E_{\rm int.} < 15.0$ MeV,
and plot their transition amplitudes as a function of $\eta_Y$ in Fig.~\ref{Fig8}.
These six typical states are chosen, due to the following reasons:
The states at $E_{\rm int.}^{(i)} = 2.25$ MeV and $E_{\rm int.}^{(f)}=0.21$ MeV
are the first $1^+$states calculated from $^{128}$Te and $^{128}$Xe, respectively.
The state at $E_{\rm int.}^{(f)}=5.14$ MeV is with the largest GT$^+$ transition amplitude in the 0-9.0 MeV region.
For this state, the largest overlap factor $\langle 1_{n_f}^+ | 1_{n_i}^+ \rangle$
is produced by the state at $E_{\rm int.}^{(i)} =6.17$ MeV.
The state at $E_{\rm int.}^{(f)}=9.15$ MeV is with the largest GT$^+$ transition amplitude
(absolute value), among the states whose GT$^+$ transition amplitudes
have changed their signs in the 9.0-15.0 MeV region.
For this state, the largest overlap factor $\langle 1_{n_f}^+|1_{n_i}^+\rangle$
produced by the state lying in 9.0-15.0 MeV is the state at $E_{\rm int.}^{(i)} =10.39$ MeV.
From Fig.~\ref{Fig8}(a), we find that the GT$^-$ transition amplitudes are almost independent of $\eta_Y$,
which is consistent with that $B({\rm GT}^{-})$ is influenced little by the IS pairing.
On the other hand, the GT$^+$ transition amplitudes, in panel (b), show a strong dependence on $\eta_Y$,
and even change their signs for the higher excited states with $E_{\rm int.}^{(f)}=9.15$ MeV.
Such a strong dependence comes from the large magnitude of
the $u_{\pi}v_{\nu}$ factor in front of $Y_{\pi\nu}$ for neutron-rich nuclei.
For the GT$^{+}$ state of $E_{\rm int.}^{(f)}=9.15$ MeV,
one can observe that after $\eta_{Y}=0.4$, the GT$^{+}$ transition amplitude crosses zero.
Therefore, the summed contribution, coming from the
$(n_i, n_f)$ intermediate states with
GT$^+$ state of $E_{\rm int.}^{(f)}=9.15$ MeV,
change their signs and become negative,
as the blue up triangle shown in Fig. \ref{Fig7}.
On the other hand, for the GT$^+$ states with energies smaller than
9.0 MeV, such as those with $E_{\rm int.}^{(f)}=0.21$ MeV or $5.14$ MeV shown in the figure,
the absolute values of their transition amplitudes keep decreasing with increasing $\eta_Y$,
which causes the first feature observed in Fig.~\ref{Fig6}.
Therefore, the variation of the GT$^+$ transition amplitudes with $\eta_Y$ for different energy regions of $E_{\rm int.}$
give rise to the two features in Fig.~\ref{Fig6} mentioned above.
These two features are different because the GT$^+$ transitions
are unblocked by pairing correlation
for low-lying states with energies smaller than 9.0 MeV,
and hence there are non-negligible transition strengths
in the case of $\eta_{Y}=0.0$.
The increase of $\eta_{Y}$ can greatly
reduce the GT$^+$ transition amplitudes in this energy region,
but is not enough to change their signs.
On the other hand, for high-lying states with energies larger than 9.0 MeV,
the GT$^+$ transitions are almost blocked and have small strengths at $\eta_{Y}=0.0$.
In this case, the sign of GT$^+$ transition amplitude can be changed easily due to the increase of $Y_{\pi\nu}$ amplitude.
At the end, we conclude that it is the enhanced ground-state correlations tuned by strong IS pairing interaction that cause the negative contributions to ${M}^{ 2\nu }_{\rm GT}$ in the energy range of 9.0-15.0 MeV,
through their influence on the GT$^+$ transition amplitudes.

\section{Summary}
\label{secConclu}
In summary, we study the $2\nu\beta\beta$-decay nuclear matrix elements $M^{2\nu}_{\rm GT}$ based on the spherical Skyrme HFB+QRPA model, for 11 nuclei with experimental data and the underlying mechanism for the fulfillment of SSD or LLD hypothesis is revealed.

In our systematic investigation of $M^{2\nu}_{\rm GT}$,
the suppression effect of the IS pairing is found.
By using the IS pairing strength determined through reproducing the
experimental data with different axial-vector coupling constants $g_A$,
we investigate the SSD and LLD hypotheses.
When $g_A=1.27$, the SSD hypothesis is fulfilled by the NMEs of
$^{48}$Ca, $^{82}$Se, $^{116}$Cd, $^{128}$Te, and $^{238}$U.
When $g_A=1.00$, the SSD hypothesis is fulfilled by the NMEs of
$^{238}$U, while the LLD hypothesis is fulfilled by the NMEs of
$^{76}$Ge, and $^{82}$Se.
The realization of SSD and LLD hypotheses for $^{76}$Ge, $^{82}$Se and $^{128}$Te are to a large extent caused by
the negative contributions in the running sums of the NMEs.
Through the comparison of the running sums of different $g_A$, we find that
by increasing $g_A$, or alternatively $f_{\rm IS}$, the negative contributions can either be induced or enlarged.

We pick up $^{128}$Te as an example to further study the reason for these negative contributions in the running sum.
By comparing the running sums of QRPA and QTDA models,
the negative contributions are found related with the strong ground-state correlations induced by large IS pairing strength,
which shows the importance of QRPA calculations over the simpler QTDA calculations for NME of $2\nu\beta\beta$ decay.
Through the study of GT$^-$ and GT$^+$ transition strength functions,
it is shown that the ground-state correlations influence the NME through its crucial role in GT$^+$ transitions
for these double-$\beta$ decay nuclei with neutron excess, where many GT$^+$ transition channels are blocked because of Pauli principle.

With the inclusion of ground-state correlations in QRPA calculations,
the increase of IS pairing strength will make their contributions larger
so that they lead to the suppression of NME contributed from the energy region of intermediate states $E_{\rm int.}< 9.0$ MeV
as well as  negative contributions from $9.0 < E_{\rm int.}< 15.0$ MeV.
Ground-state correlations  play their roles on the above two features
by suppressing  GT$^{+}$ transition amplitudes of low-lying states
and changing the signs of GT$^{+}$ transition amplitudes of higher-lying states.

\section*{Acknowledgement}
Y. F. Niu and W. L. Lv acknowledge the support of the National Natural Science
Foundation of China under Grant No. 12075104, and the Fundamental
Research Funds for the Central Universities under
Grants No. lzujbky-2021-it10.
D. L. Fang acknowledges the support of  the ``Light of West'' program and the ``from zero to one" program by  CAS.
C. L. Bai acknowledges the support of the National Natural Science
Foundation of China under Grants No. 11575120 and 11822504.

\end{document}